\documentclass[11pt]{article}

\usepackage{graphicx,amsmath,amsfonts,amssymb, dcolumn,amsthm}
\usepackage{slashed}
\usepackage{bbm}                
\usepackage{xspace}
\usepackage{pgffor}	
\usepackage{tikz}
\usepackage{mathbbol}
\usepackage{latexsym}
\usepackage{bbm} 
\usepackage{slashed}            
\usepackage{bbm}                
\usepackage{xspace}				
\usepackage[colorlinks,citecolor=blue,urlcolor=blue,linkcolor=blue]{hyperref}

\numberwithin{equation}{section} 
\begin{document}

\title{\textbf{Equivalence between the Lovelock-Cartan action and a constrained gauge theory}}

\author{\textbf{O.~C.~Junqueira$^1$}\thanks{octavio@if.uff.br}\ , \textbf{A.~D.~Pereira$^2$}\thanks{duarte763@gmail.com}\ , \textbf{G.~Sadovski$^1$}\thanks{gsadovski@if.uff.br}\ , \textbf{T.~R.~S.~Santos$^1$}\thanks{tiagoribeiro@if.uff.br}\ , \\ \textbf{R.~F.~Sobreiro$^1$}\thanks{sobreiro@if.uff.br}\ , \textbf{A.~A.~Tomaz$^{1,3}$}\thanks{tomaz@cbpf.br}\\\\
\textit{{\small $^1$UFF - Universidade Federal Fluminense, Instituto de F\'isica,}}\\
\textit{{\small Campus da Praia Vermelha, Av. General Milton Tavares de Souza s/n, 24210-346,}}\\
\textit{{\small Niter\'oi, RJ, Brasil.}}\\
\textit{{\small $^2$UERJ - Universidade Estadual do Rio de Janeiro, Departamento de F\'isica Te\'orica,}}\\
\textit{{\small Rua S\~ao Francisco Xavier 524, 20550-013, Maracan\~a,}}\\
\textit{{\small Rio de Janeiro, RJ, Brasil.}}\\
\textit{{\small $^3$CBPF - Centro Brasileiro de Pesquisas F\'isicas,}}\\
\textit{{\small Rua Dr. Xavier Sigaud 150, 22290-180,}}\\
\textit{{\small Rio de Janeiro, RJ, Brasil.}}}
\date{}
\maketitle

\begin{abstract}
We show that the four-dimensional Lovelock-Cartan action can be derived from a massless gauge theory for the $SO(1,3)$ group with an additional BRST trivial part. The model is originally composed by a topological sector and a BRST exact piece and has no explicit dependence on the metric, the vierbein or a mass parameter. The vierbein is introduced together with a mass parameter through some BRST trivial constraints. The effect of the constraints is to identify the vierbein with some of the additional fields, transforming the original action into the Lovelock-Cartan one. In this scenario, the mass parameter is identified with Newton's constant while the gauge field is identified with the spin-connection. The symmetries of the model are also explored. Moreover, the extension of the model to a quantum version is qualitatively discussed. 
\end{abstract}

\section{Introduction}
\label{INTRO}

In \cite{Mardones:1990qc}, J.~Zanelli and A.~Mardones proposed the most general gravity action depending on the curvature and torsion without the use of the Hodge dual operation for any spacetime dimension. This result generalizes the Lovelock theorem \cite{Lovelock:1971yv} which states, for any dimension, the most general gravity action depending only on the curvature. The Zanelli-Mardones result was baptized as \emph{Lovelock-Cartan theory of gravity}. The main motivations of this result, in despite of the fact that torsion degrees of freedom have never been observed in gravity, is that torsion might be relevant at quantum level and the fact that curvature and torsion are at the same level under the geometry point of view \cite{Mardones:1990qc,Trautman:2006fp,cartan:1922,cartan:1923,cartan:1924,cartan:1925,DeSabbata:1986sv,Hehl:1976kj}.

The fact that curvature and torsion are independent quantities in Lovelock-Cartan (LC) theories enables the use of the Einstein-Cartan formalism of gravity \cite{Trautman:2006fp,Utiyama:1956sy,Kibble:1961ba,Sciama:1964wt}, which is based on the \emph{vierbein} and \emph{spin-connection} as fundamental and independent variables. In this approach, gravity can be interpreted as a kind of gauge theory where the gauge symmetry is identified with the spacetime local isometries. This equivalence opens the possibility of the application of well-known quantization techniques of gauge theories.

In despite of how similar such formulations of gravity are with respect to gauge theories, their quantization as fundamental theories still lacks. Essentially, these theories share the same problems of pure metric and Palatini theories of gravity \cite{Ashtekar:2014ife,Rovelli:1997qj,Baez:1995sj}. In particular, the perturbative renormalizability or unitarity problems remain \cite{Buchbinder:1992rb}, as well as background independence \cite{Ashtekar:2014kba,Rovelli:2014ssa} and so on.


In the present work we provide the construction of a gauge theory which encodes the Lovelock-Cartan dynamics. The gauge theory is constructed for the gauge group $SO(1,3)$ over a four-dimensional manifold. In contrast to the gravity theories in the Einstein-Cartan formalism, the gauge degrees of freedom and the spacetime are independent, by construction. The original action is massless and is formed by a topological term and a BRST exact one and the fundamental fields are the gauge connection, the ghost field and a quartet system formed by two BRST doublets. Moreover, the gauge theory is metric independent and also independent of the vierbein field. On the side of the manifold, we provide no dynamics to it. It is just a generic manifold where the gauge theory lives on. Hence, with the help of extra BRST doublets, we introduce an algebraic quadratic coupling with the vierbein of the manifold. The extra doublets can be visualized as Lagrange multipliers for extra constraints. The effect of such constraints is to transform the gauge theory coupled to the vierbein into the four-dimensional LC action. Essentially, the constraints identify the gauge theory degrees of freedom with spacetime, providing the LC dynamics to it.

The model enjoys a rich set of symmetries that can be written as consistent Ward identities. This feature would be important in a quantum version of the model. A possibility is to quantize the gauge theory coupled to the classical vierbein. The classical limit of such model would be the LC action. In this scenario, the dynamics of spacetime would be ruled by a quantum gauge theory composed by a topological piece and a BRST exact one. Nevertheless, in this work, we remain at classical level. The formalization of the quantum version of the model is left for future investigation due to the intricacies of renormalizability and gauge fixing of metric free theories.

The article is organized as follows: In Sec.~\ref{EFGRAV} we provide a small overview of the Lovelock-Cartan action in four dimensions. In Sec.~\ref{TGT} we construct the massless gauge theory composed by a topological term and a BRST exact one. We also provide a complete discussion about the symmetries of the model in terms of Ward identities. In Sec.~\ref{CA} we introduce the massive constraint carrying the vierbein classical field and discuss how the constraint leads to the LC action. In addition, we generalize all Ward identities of the previous section. In Sec.~\ref{DIS} we provide an extra discussion about the BRST symmetry and a detailed, yet qualitative, discussion about the quantum version of the model. Finally, in Sec.~\ref{CONC} we display our conclusions.

\section{Overview of the Lovelock-Cartan action in four dimensions}\label{EFGRAV}

The Lovelock-Cartan action \cite{Mardones:1990qc} in four dimensions is given by
\begin{equation}
S=S_0+S_\mu\;,\label{action1}
\end{equation}
where
\begin{equation}
 S_0=\int \left(z_1\epsilon_{abcd}R^{ab}R^{cd}+z_2R^{ab}R_{ab}\right)\;,\label{action1a}
 \end{equation}
and
\begin{equation} 
  S_\mu=\mu^2\int\left[\epsilon_{abcd}\left(z_3R^{ab}e^ce^d+z_4\mu^2e^ae^be^ce^d\right)+z_5R_{ab}e^ae^b+z_6T^aT_a\right]\;,\label{action1b}
\end{equation}
over a four-dimensional manifold $M$. The quantities $R^a_{\phantom{a}b}=\mathrm{d}\omega^a_{\phantom{a}b}+\omega^a_{\phantom{a}c} \omega^c_{\phantom{c}b}$ and $T^a=De^a=\mathrm{d}e^a+\omega^a_{\phantom{a}b}e^b$ are, respectively, the curvature and torsion 2-forms. The basic fields are the vierbein 1-form $e^a$ and the connection 1-form $\omega^a_{\phantom{a}b}$. All parameters $z_i$ are dimensionless while $\mu$ carries mass dimension 1. Moreover, the action \eqref{action1} also contains the invariant tensors $\epsilon_{abcd}$ and $\eta_{ab}$.

The first term in \eqref{action1a} is recognized as the Gauss-Bonnet topological term while the second term is the Pontryagin topological term. For $z_6=-z_5$, the last two terms in \eqref{action1b} are also reduced to a topological term, \emph{i.e.}, the Nieh-Yan term: $T^aT_a-R_{ab}e^ae^b=d(e_aT^a)$. On the other hand, because $DT=Re$, these terms are actually the same up to surface terms. Thus, generically, $S_0$ is topological while $S_\mu$ is dynamical. Obviously, the first term in the action $S_\mu$ is the Einstein-Hilbert action while the second term in $S_\mu$ is the a cosmological constant term. Hence, $\mu^2z_3$ is identified with Newton's constant while $\mu^2z_4/z_3$  with the cosmological constant. 

The action \eqref{action1} is invariant under gauge transformations for the group $SO(1,3)$ whose infinitesimal version are 
\begin{eqnarray}
\delta e^a&=&-\xi^a_{\phantom{a}b}e^b\;,\nonumber\\
\delta\omega^a_{\phantom{a}b}&=&D\xi^a_{\phantom{a}b}\;=\;\textrm{d}\xi^a_{\phantom{a}b}+\omega^a_{\phantom{a}c}\xi^c_{\phantom{a}b}-
\omega_b^{\phantom{b}c}\xi_c^{\phantom{c}a}\;,\label{gt1}
\end{eqnarray}
which describe the local spacetime isometries associated to the strong equivalence principle. The quantity $\xi^a_{\phantom{a}b}$ is the infinitesimal gauge parameter and $D$ the exterior covariant derivative in the adjoint representation. These transformations correspond to transformations in the cotangent space at a point $x\in M$. Moreover, since these transformations leave the manifold coordinates unchanged, they can be interpreted as gauge transformations. Thus, gravity can be interpreted as a special type of gauge theories for which the fields have a geometrical meaning\footnote{The fields directly determine the spacetime dynamics.}. These transformation laws establish that, under the gauge theory point of view, the gauge field of the model is the spin-connection $\omega$ while the vierbein $e$ is a matter field.

\section{A massless gauge theory}\label{TGT}

As discussed at the Introduction, the aim of the paper is to show that the LC action \eqref{action1} can be obtained from a trivial theory (in the sense of containing just a topological and BRST-exact terms) by the introduction of a suitable algebraic linear constraint. This section is devoted to the construction of such trivial action.

\subsection{Fundamental ingredients and action}

We consider a massless $SO(1,3)$ gauge theory in a four-dimensional manifold. The natural ingredients are the fundamental fields of the theory, namely the gauge field $\omega^a_{\phantom{a}b}$ and the ghost field $c^a_{\phantom{a}b}$, and the invariant tensors $\epsilon_{abcd}$ and $\eta_{ab}$. The most general action with vanishing ghost number, dimension four, polynomial on the fields and their derivatives and not explicitly dependent on the metric or the vierbein is the topological part of the LC action, namely $S_0$. Inhere, for consistency, we define it as
\begin{equation}
S_0^\prime=\int \left(z^\prime_1\epsilon_{abcd}R^{ab}R^{cd}+z^\prime_2R^{ab}R_{ab}\right)\;,\label{action1aa}
\end{equation}
where $z_i^\prime$ are dimensionaless parameters which, eventually will be identified with the original parameters $z_i$ of the topological action \eqref{action1a}. Because the model is massless by construction, there is no room for $S_\mu$-like terms. Hence, the vierbein independence is ensured at this point. The fundamental fields transform under BRST symmetry as
\begin{eqnarray}
s\omega^a_{\phantom{a}b}&=&-Dc^a_{\phantom{a}b}\;,\nonumber\\
sc^a_{\phantom{a}b}&=&-c^a_{\phantom{a}c}c^c_{\phantom{c}b}\;,\label{brst1}
\end{eqnarray}
where $s$ is the nilpotent BRST operator.

We also define a BRST quartet system of dimensionless 1-forms, namely,
\begin{eqnarray}
s\bar{\eta}^a&=&\bar{\sigma}^a-c^a_{\phantom{a}b}\bar{\eta}^b\;,\nonumber\\
s\bar{\sigma}^a&=&-c^a_{\phantom{a}b}\bar{\sigma}^b\;,\nonumber\\
s\sigma^a&=&\eta^a-c^a_{\phantom{a}b}\sigma^b\;,\nonumber\\
s\eta^a&=&-c^a_{\phantom{a}b}\eta^b\;.\label{brs2}
\end{eqnarray}
The 1-forms $\bar{\eta}^a$ and $\eta^a$ have fermionic statistics while $\bar{\sigma}^a$ and $\sigma^a$ have a bosonic one. Moreover, it is clear from the transformations \eqref{brs2} that the quartet is a double BRST doublet. This means that they are non-physical fields, belonging to the trivial sector of the BRST cohomology \cite{Piguet:1995er}. The existence of the quartet system allows the introduction of an extra term to the action 
\begin{eqnarray} 
S_{triv}&=&s\int\left\{\epsilon_{abcd}\left(z^{\prime}_3R^{ab}\bar{\eta}^c\sigma^d+z^{\prime}_4\bar{\eta}^a\sigma^b\bar{\sigma}^c\sigma^d
\right)+z^{\prime}_5R_{ab}\bar{\eta}^a\sigma^b+z^{\prime}_6D\bar{\eta}^aD\sigma_a+\right.\nonumber\\
&+&\left.z^{\prime}_7\left[\bar{\eta}^a\sigma_a\left(\bar{\sigma}^b\sigma_b-\bar{\eta}^b\eta_b\right)+
\bar{\eta}^a\sigma_b\bar{\eta}_a\eta^b\right]\right\}\nonumber\\
&=&\int\left\{\epsilon_{abcd}\left[z^{\prime}_3R^{ab}\left(\bar{\sigma}^c\sigma^d+\bar{\eta}^c\eta^d\right)+
z^{\prime}_4\left(\bar{\sigma}^a\sigma^b+\bar{\eta}^a\eta^b\right)\left(\bar{\sigma}^c\sigma^d+\bar{\eta}^c\eta^d\right)\right]+
\right.\nonumber\\
&+&\left.z^{\prime}_5R_{ab}\left(\bar{\sigma}^a\sigma^b+\bar{\eta}^a\eta^b\right)-z^{\prime}_6\left(D\bar{\sigma}^aD\sigma_a-D\bar{\eta}^aD\eta_a\right)+\right.\nonumber\\
&+&\left.z^{\prime}_7\left[\left(\bar{\sigma}^a\sigma_a+\bar{\eta}^a\eta_a\right)\left(\bar{\sigma}^b\sigma_b-\bar{\eta}^b\eta_b\right)
+2\bar{\eta}_a\eta^b\left(\bar{\sigma}^a\sigma_b-
\bar{\sigma}_b\sigma^a\right)+\bar{\eta}^a\eta^b\bar{\eta}_a\eta_b\right]\right\}\;, \label{actiontriv}
\end{eqnarray}
which is trivial with respect to the BRST cohomology. The parameters $z_i^{\prime}$ are dimensionless parameters\footnote{Eventually, they will be associated with the Lovelock-cartan parameters $z_i$ appearing in \eqref{action1}.}.

The action 
\begin{equation}
S_T=S^\prime_0+S_{triv}\;,\label{actionT}
\end{equation}
is dynamically empty because the topological term does not contribute to the field equations and the BRST exact sector ensures that $S_{triv}$ is dynamically trivial. Moreover, the action \eqref{actionT} is highly non-perturbative\footnote{It is not difficult to check that the action \eqref{actionT}, as it stands, has no quadratic terms in the fields. As a consequence, there is no free theory to be defined (and no tree-level propagators). Hence, a perturbative expansion around a free theory is not at our disposal. In fact, all non-vanishing terms in \eqref{actionT} are interacting terms. A theory of this type is said to be highly non-perturbative. Of course, one can always define background configurations and enforce a perturbative regime around these configurations.} due to the absence of quadratic terms. We also remark that this action is independent of the metric, the vierbein and the ghost fields. As a consequence the independence of the vierbein field in the action can be taken as an independence between the gauge symmetry and the manifold isometries. This property ensures that the dynamics of spacetime is not related to any of the fields in \eqref{actionT}.

\subsection{Symmetries and Ward identities}

All continuous symmetries of the action $S_T$ can be characterized in a functional way through suitable Ward identities. It is useful to define a set of BRST invariant sources in order to control the non-linear character of the BRST transformations of the fields through the external action
\begin{eqnarray}
S_{ext}&=&s\int\left(\Omega_a^{\phantom{a}b}\omega^a_{\phantom{a}b}+L_a^{\phantom{a}b}c^a_{\phantom{a}b}-X_a\bar{\eta}^a-\bar{X}_a\eta^a+Y_a\bar{\sigma}^a+\bar{Y}_a\sigma^a\right)\nonumber\\
&=&\int\left[-\Omega_a^{\phantom{a}b}Dc^a_{\phantom{a}b}-L_a^{\phantom{a}b}c^a_{\phantom{a}c}c^c_{\phantom{c}b}+X_a\left(\bar{\sigma}^a-c^a_{\phantom{a}b}\bar{\eta}^b
\right)-\bar{X}_ac^a_{\phantom{a}b}\eta^b-Y_ac^a_{\phantom{a}b}\bar{\sigma}^b+\right.\nonumber\\
&+&\left.\bar{Y}_a\left(\eta^a-c^a_{\phantom{a}b}\sigma^b\right)\right]\;, \label{actionext}
\end{eqnarray}
with 
\begin{equation}
s\Omega_a^{\phantom{a}b}=sL_a^{\phantom{a}b}=sX_a=s\bar{X}_a=sY_a=s\bar{Y}_a=0\;.\label{brst3}
\end{equation}
The full action is then
\begin{equation}
\Sigma_0=S_T+S_{ext}\;.
\end{equation}

We now can list all Ward identities.
\begin{itemize}
\item Slavnov-Taylor identity:
\begin{equation}
\mathcal{S}(\Sigma_0)=0\;,\label{STeq0}
\end{equation}
where
\begin{eqnarray}
\mathcal{S}(\Sigma_0)&=&\int\left(\frac{\delta\Sigma_0}{\delta\Omega_a^{\phantom{a}b}}\frac{\delta\Sigma_0}{\delta\omega^a_{\phantom{a}b}}+\frac{\delta\Sigma_0}{\delta L_a^{\phantom{a}b}}\frac{\delta\Sigma_0}{\delta c^a_{\phantom{a}b}}+\frac{\delta\Sigma_0}{\delta X_a}\frac{\delta\Sigma_0}{\delta\bar{\eta}^a}+\frac{\delta\Sigma_0}{\delta\bar{X}_a}\frac{\delta\Sigma_0}{\delta\eta^a}+\frac{\delta\Sigma_0}{\delta Y_a}\frac{\delta\Sigma_0}{\delta\bar{\sigma}^a}+\right.\nonumber\\
&+&\left.\frac{\delta\Sigma_0}{\delta\bar{Y}_a}\frac{\delta\Sigma_0}{\delta\sigma^a}\right)\;.
\label{STop0}
\end{eqnarray}

\item Ghost equation:
\begin{equation}
\int\frac{\delta\Sigma_0}{\delta c^a_{\phantom{a}b}}=\Delta_a^{\phantom{a}b}\;,\label{Ghosteq0}
\end{equation}
where
\begin{equation}
\Delta_a^{\phantom{a}b}=\int\left(-L_a^{\phantom{a}c}c^b_{\phantom{b}c}+L_c^{\phantom{c}b}c^c_{\phantom{c}a}-\Omega_a^{\phantom{a}c}\omega^b_{\phantom{b}c}+
\Omega_c^{\phantom{c}b}\omega^c_{\phantom{c}a}+X_a\bar{\eta}^b+\bar{X}_a\eta^b-Y_a\bar{\sigma}^b-\bar{Y}_a\sigma^b\right)\;,\label{brk0}
\end{equation}
is a linear breaking.

\item Vierbein equation:
\begin{equation}
\frac{\delta\Sigma_0}{\delta e^a}=0\;.\label{vierbein0}
\end{equation}

\item Rigid supersymmetries:
\begin{equation}
R^{(i)}\Sigma_0=\Delta^{(i)}\;,\label{super0}
\end{equation}
where $i\in\{1,2,3,4\}$. The rigid supersymmetric operators are:
\begin{eqnarray}
R^{(1)}&=&\sigma^a\frac{\delta}{\delta\eta^a}-\bar{\eta}^a\frac{\delta}{\delta\bar{\sigma}^a}-Y^a\frac{\delta}{\delta X^a}-\bar{X}^a\frac{\delta}{\delta\bar{Y}^a}\;,\nonumber\\
R^{(2)}&=&\bar{\sigma}^a\frac{\delta}{\delta\bar{\eta}^a}+\eta^a\frac{\delta}{\delta\sigma^a}-X^a\frac{\delta}{\delta Y^a}+\bar{Y}^a\frac{\delta}{\delta\bar{X}^a}\;,\nonumber\\
R^{(3)}&=&\bar{\sigma}^a\frac{\delta}{\delta\eta^a}-\bar{\eta}^a\frac{\delta}{\delta\sigma^a}-\bar{Y}^a\frac{\delta}{\delta X^a}-\bar{X}^a\frac{\delta}{\delta Y^a}\;,\nonumber\\
R^{(4)}&=&\sigma^a\frac{\delta}{\delta\bar{\eta}^a}+\eta^a\frac{\delta}{\delta\bar{\sigma}^a}+Y^a\frac{\delta}{\delta\bar{X}^a}-X^a\frac{\delta}{\delta\bar{Y}^a}\;,\label{super0a}
\end{eqnarray}
while the only non-vanishing $\Delta^{(i)}$ are
\begin{eqnarray}
\Delta^{(1)}&=&X_a\bar{\eta}^a-\bar{X}_a\eta^a-Y_a\bar{\sigma}^a+\bar{Y}_a\sigma^a\;,\nonumber\\
\Delta^{(4)}&=&-2X_a\eta^a\;,\label{super0b}
\end{eqnarray}
which are linear in the fields.

\item Rigid fermionic equations:
\begin{equation}
Q^{(i)}\Sigma_0=\Lambda^{(i)}\;,\label{fermionic0}
\end{equation}
where
\begin{eqnarray}
Q^{(1)}&=&-\bar{\eta}^a\frac{\delta}{\delta\eta^a}+\bar{X}^a\frac{\delta}{\delta X^a}\;,\nonumber\\
Q^{(2)}&=&\eta^a\frac{\delta}{\delta\bar{\eta}^a}-X^a\frac{\delta}{\delta\bar{X}^a}\;,\label{fermionic0a}
\end{eqnarray}
and the only nonvanishing breaking is
\begin{equation}
\Lambda^{(1)}=\bar{X}_a\bar{\sigma}^a-\bar{Y}_a\bar{\eta}^a\;,\label{fermionic0b}
\end{equation}
which is linear in the fields.

\item First $U^{4}(1)$ charge equation:
\begin{equation}
Q_0\Sigma_0=0\;,\label{Qcharge0}
\end{equation}
where
\begin{eqnarray}
Q_0&=&\sigma^a\frac{\delta}{\delta\sigma^a}-\bar{\sigma}^a\frac{\delta}{\delta\bar{\sigma}^a}+\eta^a\frac{\delta}{\delta\eta^a}-\bar{\eta}^a\frac{\delta}{\delta\bar{\eta}^a}+X^a\frac{\delta}{\delta X^a}-\bar{X}^a\frac{\delta}{\delta\bar{X}^a}+\nonumber\\
&+&Y^a\frac{\delta}{\delta Y^a}-\bar{Y}^a\frac{\delta}{\delta \bar{Y}^a}\;.
\label{Qcharge0a}
\end{eqnarray}
Equation \eqref{Qcharge0} expresses the existence of a quantum number associated with a $U^4(1)$ symmetry among the quartet fields.

\item Second $U^{4}(1)$ charge equation:
\begin{equation}
\bar{Q}_0\Sigma_0=-2(X_a\bar{\sigma}^a+\bar{Y}_a\eta^a)\;,\label{Qcharge0bar}
\end{equation}
where
\begin{eqnarray}
\bar{Q}_0&=&\sigma^a\frac{\delta}{\delta\sigma^a}-\bar{\sigma}^a\frac{\delta}{\delta\bar{\sigma}^a}-\eta^a\frac{\delta}{\delta\eta^a}+\bar{\eta}^a\frac{\delta}{\delta\bar{\eta}^a}-X^a\frac{\delta}{\delta X^a}+\bar{X}^a\frac{\delta}{\delta\bar{X}^a}+\nonumber\\
&+&Y^a\frac{\delta}{\delta Y^a}-\bar{Y}^a\frac{\delta}{\delta \bar{Y}^a}\;.
\end{eqnarray}
Equation \eqref{Qcharge0bar} expresses the existence of a second quantum number associated with the other $U^4(1)$ symmetry among the quartet fields. Combination of \eqref{Qcharge0} and \eqref{Qcharge0bar} results in
\begin{equation}
Q_{eff}^{(i)}\Sigma_0=(-1)^{i-1}(X_a\bar{\sigma}^a+\bar{Y}_a\eta^a)\;,
\end{equation}
where
\begin{eqnarray}
Q_{eff}^{(1)}&=&\frac{1}{2}\left(Q_0-\bar{Q}_0\right)=\eta^a\frac{\delta}{\delta\eta^a}-\bar{\eta}^a\frac{\delta}{\delta\bar{\eta}^a}+X^a\frac{\delta}{\delta X^a}-\bar{X}^a\frac{\delta}{\delta\bar{X}^a}\;,\nonumber\\
Q_{eff}^{(2)}&=&Q_{eff}^{(1)}+\bar{Q}_0=\sigma^a\frac{\delta}{\delta\sigma^a}-\bar{\sigma}^a\frac{\delta}{\delta\bar{\sigma}^a}+Y^a\frac{\delta}{\delta Y^a}-\bar{Y}^a\frac{\delta}{\delta\bar{Y}^a}\;.
\end{eqnarray}
\end{itemize}
For completeness, we display the quantum numbers of the fields (including the vierbein) in Table \ref{table1} and the quantum numbers of the sources in Table \ref{table2}.

The commutation relations between the Ward operators can also be obtained by a straightforward computation. Starting with the Slavnov-Taylor operator, for instance, let $\mathcal{F}$ be general functional of even ghost and form number, we define the Slavnov-Taylor operator action on $\mathcal{F}$ as 
\begin{eqnarray}
\mathcal{S}(\mathcal{F})&=&\int\left(\frac{\delta\mathcal{F}}{\delta\Omega_a^{\phantom{a}b}}\frac{\delta\mathcal{F}}{\delta\omega^a_{\phantom{a}b}}+\frac{\delta\mathcal{F}}{\delta L_a^{\phantom{a}b}}\frac{\delta\mathcal{F}}{\delta c^a_{\phantom{a}b}}+\frac{\delta\mathcal{F}}{\delta X_a}\frac{\delta\mathcal{F}}{\delta\bar{\eta}^a}+\frac{\delta\mathcal{F}}{\delta\bar{X}_a}\frac{\delta\mathcal{F}}{\delta\eta^a}+\frac{\delta\mathcal{F}}{\delta Y_a}\frac{\delta\mathcal{F}}{\delta\bar{\sigma}^a}+\frac{\delta\mathcal{F}}{\delta\bar{Y}_a}\frac{\delta\mathcal{F}}{\delta\sigma^a}\right)\;.\nonumber\\
\label{STF}
\end{eqnarray}
Its linearized version reads
\begin{eqnarray}
\mathcal{S}_{\mathcal{F}}&=&\int\left(\frac{\delta\mathcal{F}}{\delta\Omega_a^{\phantom{a}b}}\frac{\delta}{\delta\omega^a_{\phantom{a}b}}+\frac{\delta\mathcal{F}}{\delta\omega^a_{\phantom{a}b}}\frac{\delta}{\delta\Omega_a^{\phantom{a}b}}+\frac{\delta\mathcal{F}}{\delta L_a^{\phantom{a}b}}\frac{\delta}{\delta c^a_{\phantom{a}b}}+\frac{\delta\mathcal{F}}{\delta c^a_{\phantom{a}b}}\frac{\delta}{\delta L_a^{\phantom{a}b}}+\frac{\delta\mathcal{F}}{\delta X_a}\frac{\delta}{\delta\bar{\eta}^a}+\frac{\delta\mathcal{F}}{\delta\bar{\eta}^a}\frac{\delta}{\delta X_a}+\right.\nonumber\\
&+&\left.\frac{\delta\mathcal{F}}{\delta\bar{X}_a}\frac{\delta}{\delta\eta^a}+\frac{\delta\mathcal{F}}{\delta\eta^a}\frac{\delta}{\delta\bar{X}_a}+\frac{\delta\mathcal{F}}{\delta Y_a}\frac{\delta}{\delta\bar{\sigma}^a}+\frac{\delta\mathcal{F}}{\delta\bar{\sigma}^a}\frac{\delta}{\delta Y_a}+\frac{\delta\mathcal{F}}{\delta\bar{Y}_a}\frac{\delta}{\delta\sigma^a}+\frac{\delta\mathcal{F}}{\delta\sigma^a}\frac{\delta}{\delta\bar{Y}_a}\right)\;.
\label{STFL}
\end{eqnarray}
From \eqref{super0a},\; \eqref{STF} and \eqref{STFL} we get
\begin{eqnarray}
R^{(i)}\mathcal{S}(\mathcal{F})+\mathcal{S}_{\mathcal{F}}R^{(i)}(\mathcal{F})&=&0\;.
\label{commprima}
\end{eqnarray}
We also have the following commutation relations
\begin{align}
\left(R^{(i)}\right)^2&=0\;,&\left\{R^{(1)},R^{(2)}\right\}&=Q_0\;,&\left\{R^{(1)},R^{(3)}\right\}&=2Q^{(1)}\;,&\nonumber\\
\left\{R^{(1)},R^{(4)}\right\}&=0\;,&\left\{R^{(2)},R^{(3)}\right\}&=0\;,&\left\{R^{(2)},R^{(4)}\right\}&=2Q^{(2)}\;,&\nonumber\\
\left\{R^{(3)},R^{(4)}\right\}&=-\bar{Q}_0\;.\label{comm1}
\end{align}
Moreover,
\begin{align}
\left[R^{(1)},Q_0\right]&=0\;,& \left[R^{(2)},Q_0\right]&=0\;,& \left[R^{(3)},Q_0\right]&=2R^{(3)}\;,\nonumber\\
\left[R^{(4)},Q_0\right]&=-2R^{(4)}\;,& \left[R^{(1)},Q^{(1)}\right]&=0\;,& \left[R^{(1)},Q^{(2)}\right]&=R^{(4)}\;,\nonumber\\
\left[R^{(2)},Q^{(1)}\right]&=-R^{(3)}\;,&
\left[R^{(2)},Q^{(2)}\right]&=\;\;0\;,& \left[R^{(3)},Q^{(1)}\right]&=0\;,\nonumber\\
\left[R^{(3)},Q^{(2)}\right]&=R^{(2)}\;,&
\left[R^{(4)},Q^{(1)}\right]&=-R^{(1)}\;,&
\left[R^{(4)},Q^{(2)}\right]&=0\;,\nonumber\\
\left[Q^{(1)},Q^{(2)}\right]&=Q^{(1)}_{eff}\;,& \left[Q^{(i)},Q_0\right]&=(-1)^{i-1}2Q^{(i)}\;. \label{comm2}
\end{align}

The rich set of Ward identities ensures that $\Sigma_0$ is the most general local classical action, polynomial in the fields and their derivatives, with vanishing ghost number and independent on the metric and the vierbein. Hence, the Ward identities ensure the triviality of the model as well as the fact that the model has no relation with spacetime dynamics.

\begin{table}[t]
\centering
\begin{tabular}{|c|c|c|c|c|c|c|c|}
\hline
field & $\omega$ & $\bar{c}$ & $e$ & $\bar{\sigma}$ & $\sigma$ & $\bar{\eta}$ & $\eta$\\ \hline
$Q$-charge & 0 & 0 & 0 & -1 & 1 & -1 & 1 \\
$\bar{Q}$-charge & 0 & 0 & 0 & -1 & 1 & 1 & -1 \\
Ghost n$^ o$ & 0 & 1 & 0 & 0 & 0 & -1 & 1\\
Form rank & 1 & 0 & 1 & 1 & 1 & 1 & 1  \\
Statistics & 1 & 1 & 1 & 1 & 1 & 0 & 2  \\ \hline
\end{tabular}%
\caption{Quantum numbers of the fundamental fields and the quartet system.}
\label{table1}
\end{table}

\begin{table}[t]
\centering
\begin{tabular}{|c|c|c|c|c|c|c|}
\hline
source & $\Omega$ & $L$ & $\bar{X}$ & $X$ & $\bar{Y}$ & $Y$ \\ \hline
$Q$-charge & 0 & 0 & -1 & 1 & -1 & 1 \\
$\bar{Q}$-charge & 0 & 0 & 1 & -1 & -1 & 1 \\
Ghost n$^ o$ & -1 & -2 & 0 & -2 & -1 & -1 \\
Form rank & 3 & 4 & 3 & 3 & 3  & 3 \\
Statistics & 2 & 2 & 3 & 3 & 2  & 2 \\ \hline
\end{tabular}%
\caption{Quantum numbers of the sources.}
\label{table2}
\end{table}

\subsection{A remark about the BRST triviality}

The quartet system \eqref{brs2} is composed by BRST doublets, and thus, these fields live at the trivial sector of the BRST cohomology. Hence, $S_{triv}$ does not affect the physical dynamical content of the topological action $S_0$. Nevertheless, we can decompose $s$ as
\begin{equation}
s=s_o+\widetilde{\delta}\;,\quad s_o^2=\widetilde{\delta}^2=\{s_o,\widetilde{\delta}\}=0\;,\label{rel0}
\end{equation}
where $s_o$ and $\widetilde{\delta}$ act on the fields as
\begin{eqnarray}
s_o\omega^a_{\phantom{a}b}&=&-Dc^a_{\phantom{a}b}\;,\nonumber\\
s_oc^a_{\phantom{a}b}&=&-c^a_{\phantom{a}c}c^c_{\phantom{c}b}\;,\nonumber\\
s_o\bar{\eta}^a&=&-c^a_{\phantom{a}b}\bar{\eta}^b\;,\nonumber\\
s_o\bar{\sigma}^a&=&-c^a_{\phantom{a}b}\bar{\sigma}^b\;,\nonumber\\
s_o\sigma^a&=&-c^a_{\phantom{a}b}\sigma^b\;,\nonumber\\
s_o\eta^a&=&-c^a_{\phantom{a}b}\eta^b\;,\label{brst1a}
\end{eqnarray}
and
\begin{eqnarray}
\widetilde{\delta}\bar{\eta}^a&=&\bar{\sigma}^a\;,\nonumber\\
\widetilde{\delta}\bar{\sigma}^a&=&0\;,\nonumber\\
\widetilde{\delta}\sigma^a&=&\eta^a\;,\nonumber\\
\widetilde{\delta}\eta^a&=&0\;.\label{brs2a}
\end{eqnarray}
Moreover, it is easy to check that
\begin{equation}
s_oS_{triv}=\widetilde{\delta}S_{triv}=0\;,\label{inv0}
\end{equation}
and
\begin{eqnarray}
S_{triv}&\neq&s_o\;(\mathrm{something})\;,\nonumber\\
S_{triv}&=&\widetilde{\delta}\;(\mathrm{something})\;.
\end{eqnarray}
Thus, although $s$ and $\widetilde{\delta}$ define the quartet system as trivial, they are not trivial with respect to $s_o$. This means that these fields can be interpreted as physical under the $s_o$ cohomology.

\section{Introducing a massive constraint and the vierbein}\label{CA}

\subsection{Constraint action}

The LC action can be recovered from $\Sigma_0$ with the introduction of a set of suitable constraints. The only demand is that the Ward identities could be broken only by linear terms in the fields. Such requirement is essential for the extension of such Ward identities to the quantum level. In fact, it is easy to see that these constraints are given by 
\begin{align}
\bar{\sigma}^a=\sigma^a&=m e^a\;,\nonumber\\
\bar{\eta}^a=\eta^a&=0\;,\label{const1}
\end{align}
where $m$ is a mass parameter\footnote{This parameter will, eventually, be identified with $\mu$ appearing  in \eqref{action1}.} and $e^a$ the vierbein field, which is taken to be a classical field\footnote{Although we are at classical level, we mean that the vierbein would remain classical in a possible quantum scenario.}. Within this construction, the vierbein must be a BRST invariant quantity,
\begin{equation}
se^a=0\;.
\end{equation}
It is clear that these constraints introduce not only the vierbein, but also a mass scale.

To employ the constraint without spoiling the BRST triviality of $\Sigma_0$, we introduce two extra BRST quartet systems, namely,
\begin{eqnarray}
s\bar{\theta}^a&=&\bar{\lambda}^a\;,\nonumber\\
s\bar{\lambda}^a&=&0\;,\nonumber\\
s\theta^a&=&\lambda^a\;,\nonumber\\
s\lambda^a&=&0\;,\label{brs4}
\end{eqnarray}
and
\begin{eqnarray}
s\bar{\gamma}^a&=&\bar{\rho}^a\;,\nonumber\\
s\bar{\rho}^a&=&0\;,\nonumber\\
s\gamma^a&=&\rho^a\;,\nonumber\\
s\rho^a&=&0\;,\label{brs5}
\end{eqnarray}
which will work as Lagrange multipliers for the constraints \eqref{const1}. The quantum numbers of the new quartet systems are displayed at Table \ref{table3}.

\begin{table}[t]
\centering
\begin{tabular}{|c|c|c|c|c|c|c|c|c|}
\hline
field & $\bar{\theta}$ & $\theta$ & $\bar{\lambda}$ & $\lambda$ & $\bar{\gamma}$ & $\gamma$ & $\bar{\rho}$ & $\rho$\\ \hline
$Q$-charge & -1 & 1 & -1 & 1 & -1 & 1 & -1 & 1 \\
$\bar{Q}$-charge & 1 & -1 & -1 & 1 & -1 & 1 & 1 & -1 \\
Ghost n$^ o$ & -1 & -1 & 0 & 0 & -2 & 0 & -1 & 1\\
Form rank & 3 & 3 & 3 & 3 & 3 & 3 & 3 & 3  \\
Statistics & 2 & 2 & 3 & 3 & 1 & 3 & 2  & 4 \\ \hline
\end{tabular}%
\caption{Quantum numbers of the constraint fields.}
\label{table3}
\end{table}

Then, the constraint action is given by
\begin{eqnarray}
S_c&=&s\int\left[\bar{\theta}_a\left(\sigma^a-m e^a\right)+\theta_a\left(\bar{\sigma}^a-m e^a\right)+\bar{\gamma}^a\eta_a+\gamma_a\bar{\eta}^a\right]\nonumber\\
&=&\int\left[\bar{\lambda}_a\left(\sigma^a-m e^a\right)+\lambda_a\left(\bar{\sigma}^a-m e^a\right)+\bar{\theta}_a\left(\eta^a-c^a_{\phantom{a}b}\sigma^b\right)-\gamma_a\left(\bar{\sigma}^a-c^a_{\phantom{a}b}\bar{\eta}^b\right)+\right.\nonumber\\
&+&\bar{\rho}_a\eta^a+\left.\rho_a\bar{\eta}^a-\theta_ac^a_{\phantom{a}b}\bar{\sigma}^b+\bar{\gamma}_ac^a_{\phantom{a}b}\eta^b\right]\;.\label{action5}
\end{eqnarray}
The action of interest is then,
\begin{equation}
\Sigma=\Sigma_0+S_c\;,\label{action6}
\end{equation}
which is totally equivalent to the LC action \eqref{action1} if a proper relation between $\{z_i^\prime,m\}$ and $\{z_i,\mu\}$ is obeyed. The first step to check this is to set all external sources in \eqref{action6} to zero. Hence, we perform the elimination of the auxiliary fields defined in \eqref{brs4} and \eqref{brs5} by the implementation of their field equations (see equations from \eqref{FE1} to \eqref{FE2}).  This will lead to an action that is totally equivalent to the Lovelock-Cartan action \eqref{action1} if the gauge parameters $\{z_i^\prime,m\}$ and the LC parameters $\{z_i,\mu\}$ are related accordingly to
\begin{align}
{z'}_1&=z_1\;, & {z'}_2&=z_2\;,& {z'}_3m^2&=z_3\mu^2\;, \nonumber\\
{z'}_4m^4&=z_4\mu^4\;,&{z'}_5m^2&=z_5\mu^2\;,&-{z'}_6m^2&=z_6\mu^2\;.
\end{align}
On the other hand, at quantum level the set $\{z_i^\prime,m\}$ might need renormalization before being identified with the LC parameters. A discussion about the quantum scenario can be found at Sec. \ref{DISa}.

\subsection{Generalized Ward identities}

The set of Ward identities enjoyed by the action \eqref{action6} are listed here:
\begin{itemize}
\item Slavnov-Taylor identity:
\begin{equation}
\mathcal{S}(\Sigma)=0\;,\label{STeq1}
\end{equation}
where
\begin{eqnarray}
\mathcal{S}(\Sigma)&=&\int\left(\frac{\delta\Sigma}{\delta\Omega_a^{\phantom{a}b}}\frac{\delta\Sigma}{\delta\omega^a_{\phantom{a}b}}+\frac{\delta\Sigma}{\delta L_a^{\phantom{a}b}}\frac{\delta\Sigma}{\delta c^a_{\phantom{a}b}}+\frac{\delta\Sigma}{\delta X_a}\frac{\delta\Sigma}{\delta\bar{\eta}^a}+\frac{\delta\Sigma}{\delta\bar{X}_a}\frac{\delta\Sigma}{\delta\eta^a}+\frac{\delta\Sigma}{\delta Y_a}\frac{\delta\Sigma}{\delta\bar{\sigma}^a}+\frac{\delta\Sigma}{\delta\bar{Y}_a}\frac{\delta\Sigma}{\delta\sigma^a}+\right.\nonumber\\
&+&\left.\bar{\lambda}^a\frac{\delta\Sigma}{\delta\bar{\theta}^a}+\lambda^a\frac{\delta\Sigma}{\delta\theta^a}+\bar{\rho}^a\frac{\delta\Sigma}{\delta\bar{\gamma}^a}+\rho^a\frac{\delta\Sigma}{\delta\gamma^a}\right)\;.\nonumber\\
\label{STop1}
\end{eqnarray}

\item Ghost equation:
\begin{equation}
\int\left(\frac{\delta\Sigma}{\delta c^a_{\phantom{a}b}}+\bar{\theta}_a\frac{\delta\Sigma}{\delta\bar{\lambda}_b}+\theta_a\frac{\delta\Sigma}{\delta\lambda_b}+\bar{\gamma}_a\frac{\delta\Sigma}{\delta\bar{\rho}_b}+\gamma_a\frac{\delta\Sigma}{\delta\rho_b}\right)=\widetilde{\Delta}_a^{\phantom{a}b}\;,\label{Ghosteq1}
\end{equation}
where
\begin{equation}
\widetilde{\Delta}_a^{\phantom{a}b}=\Delta_a^{\phantom{a}b}-m\int\left(\bar{\theta}_a+\theta_a\right)e^b\;,\label{brk1}
\end{equation}
remains a linear breaking.

\item Vierbein equation:
\begin{equation}
\frac{\delta\Sigma}{\delta e^a}=m\left(\bar{\lambda}^a+\lambda^a\right)\;,\label{vierbein1}
\end{equation}
which is linearly broken.

\item Rigid supersymmetries:
\begin{equation}
\widetilde{R}^{(i)}\Sigma=\widetilde{\Delta}^{(i)}\;,\label{super1}
\end{equation}
where $i\in\{1,2,3,4\}$. The rigid supersymmetric operators are\footnote{The symmetry $R^{(4)}$ is quadraticaly broken, and thus, it is not an interesting identity for the model. As a consequence, generalized versions of $Q^{(2)}$ and $\bar{Q}_0$ are not at our disposal in the full model. Obviously, since there is no generalization of the $\bar{Q}_0$ symmetry, there is no place for generalizing $Q^{(i)}_{eff}$ as well.}:
\begin{eqnarray}
\widetilde{R}^{(1)}&=&R^{(1)}+\theta^a\left(\frac{\delta}{\delta\gamma^a}+\frac{\delta}{\delta\lambda^a}\right)+\bar{\gamma}^a\left(\frac{\delta}{\delta\bar{\theta}^a}-\frac{\delta}{\delta\bar{\rho}^a}\right)-\left(\bar{\theta}^a+\bar{\rho}^a\right)\frac{\delta}{\delta\bar{\lambda}^a}-\left(\lambda^a-\gamma^a\right)\frac{\delta}{\delta\rho^a}\;,\nonumber\\
\widetilde{R}^{(2)}&=&R^{(2)}+\gamma^a\frac{\delta}{\delta\theta^a}-\bar{\theta}^a\frac{\delta}{\delta\bar{\gamma}^a}+\bar{\lambda}^a\frac{\delta}{\delta\bar{\rho}^a}-\rho^a\frac{\delta}{\delta\lambda^a}\;,\nonumber\\
\widetilde{R}^{(3)}&=&R^{(3)}+\bar{\gamma}^a\frac{\delta}{\delta\theta^a}+\bar{\theta}^a\frac{\delta}{\delta\gamma^a}-\bar{\rho}^a\frac{\delta}{\delta\lambda^a}-\bar{\lambda}^a\frac{\delta}{\delta\rho^a}\;,\label{super1a}
\end{eqnarray}
while
\begin{equation}
\widetilde{\Delta}^{(i)}=\Delta^{(i)}+\Upsilon^{(i)}\;,\label{super1b}
\end{equation}
with
\begin{eqnarray}
\Upsilon^{(1)}&=&m\left(\bar{\theta}_a-\theta_a+\bar{\rho}_a\right)e^a\;,\nonumber\\
\Upsilon^{(2)}&=&m{\rho}_ae^a\;,\nonumber\\
\Upsilon^{(3)}&=&m\bar{\rho}_ae^a\;,\label{super1c}
\end{eqnarray}
which are linear in the fields.

\item Rigid fermionic equation:
\begin{equation}
\widetilde{Q}^{(1)}\Sigma=\widetilde{\Lambda}^{(1)}\;.\label{fermionic1}
\end{equation}
where
\begin{eqnarray}
\widetilde{Q}^{(1)}&=&Q^{(1)}+\left(\bar{\theta}^a+\bar{\rho}^a\right)\frac{\delta}{\delta\rho^a}+\bar{\gamma}^a\left(\frac{\delta}{\delta\gamma^a}+\frac{\delta}{\delta\lambda^a}\right)\;,\nonumber\\
\widetilde{\Lambda}^{(1)}&=&\Lambda^{(1)}-m\bar{\gamma}_ae^a\;.\label{fermionic1a}
\end{eqnarray}
but is also linear in the fields.

\item $U^{4}(1)$ charge equation:
\begin{equation}
\widetilde{Q}_0\Sigma= m\left(\bar{\lambda}_a-\lambda_a\right)e^a\;,\label{Qcharge1}
\end{equation}
where
\begin{eqnarray}
\widetilde{Q}_0&=&\sigma^a\frac{\delta}{\delta\sigma^a}-\bar{\sigma}^a\frac{\delta}{\delta\bar{\sigma}^a}+\eta^a\frac{\delta}{\delta\eta^a}-\bar{\eta}^a\frac{\delta}{\delta\bar{\eta}^a}+X^a\frac{\delta}{\delta X^a}-\bar{X}^a\frac{\delta}{\delta\bar{X}^a}+Y^a\frac{\delta}{\delta Y^a}-\bar{Y}^a\frac{\delta}{\delta Y^a}+\nonumber\\
&+&\lambda^a\frac{\delta}{\delta\lambda^a}-\bar{\lambda}^a\frac{\delta}{\delta\bar{\lambda}^a}+\theta^a\frac{\delta}{\delta\theta^a}-\bar{\theta}^a\frac{\delta}{\delta\bar{\theta}^a}+\rho^a\frac{\delta}{\delta\rho^a}-\bar{\rho}^a\frac{\delta}{\delta\bar{\rho}^a}+\gamma^a\frac{\delta}{\delta\gamma^a}-\bar{\gamma}^a\frac{\delta}{\delta\gamma^a}\;.
\label{Qcharge1a}
\end{eqnarray}
Equation \eqref{Qcharge1} still expresses the existence of a quantum number associated with a $U^4(1)$ symmetry among the quartet fields, even though the symmetry is linearly broken.

\item Field equations:
\begin{eqnarray}
\frac{\delta\Sigma}{\delta\bar{\lambda}_a}&=&\sigma^a-m e^a\;, \label{FE1}\\
\frac{\delta\Sigma}{\delta\lambda_a}&=&\bar{\sigma}^a-m e^a\;,\\
\frac{\delta\Sigma}{\delta\bar{\theta}_a}-\frac{\delta\Sigma}{\delta\bar{Y}_a}&=&0\;,\\
\frac{\delta\Sigma}{\delta\theta_a}-\frac{\delta\Sigma}{\delta Y_a}&=&0\;,\\
\frac{\delta\Sigma}{\delta\bar{\gamma}_a}+\frac{\delta\Sigma}{\delta\bar{X}_a}&=&0\;,\\
\frac{\delta\Sigma}{\delta\gamma_a}+\frac{\delta\Sigma}{\delta X_a}&=&0\;,\\
\frac{\delta\Sigma}{\delta\rho_a}&=&\eta^a\;,\\
\frac{\delta\Sigma}{\delta\rho_a}&=&\bar{\eta}^a\;.\label{FE2}
\end{eqnarray}
\end{itemize}

We can conclude at this point that the effect of the introduction of the constraints is that most of the Ward identities are linearly broken. Moreover, the relations \eqref{commprima}, \eqref{comm1} and \eqref{comm2} are easily generalized by rejecting all relations of $R^{(4)}$, $Q^{(2)}$ and $\bar{Q}_0$.

\section{Discussion}\label{DIS}

\subsection{Quantization attempts}\label{DISa}

In light of gauge theories, let us take a closer look at the action \eqref{action1}. First of all, we see that there are no quadratic terms\footnote{This is a problem that also appear at the pure Einstein-Hilbert action.} in \eqref{action1}, only interacting terms in the fields $\omega$ and $e$. This is a problem if one plans to quantize the LC action because this property ruins the well established perturbative program of QFT, unless a background is previously chosen. Background independence though requires that such choice is arbitrary.

Another problem to be faced is the gauge fixing. A typical gauge fixing is obtained by fixing the divergence of the gauge field. However, to define the divergence of a field, the Hodge dual operator is required. Hence, an explicit dependence on the metric should be introduced.

The advantage in working with the action $\Sigma$ instead is that the model can be interpreted as a typical gauge theory for the gauge field $\omega$ and the fields defined in \eqref{brs2}. Hence, the theory is composed by a topological piece and a BRST trivial sector. The addition of the constraint \eqref{actionT} introduces a coupling with the vierbein in such a way that (some of) the BRST trivial fields are identified with the vierbein. Thus, the interpretation of the field $\omega$ as the spin-connection is natural.

A quantum version of such model would also be highly non-perturbative since there are no quadratic terms of $\omega$ in $\Sigma$. Moreover, the terms in the constraint action $S_c$ are algebraic, i.e., there are no kinetic terms. To face this problem one should, perhaps, employ the strategies developed in \cite{Baulieu:1988xs}. The authors in \cite{Baulieu:1988xs} claim that a BRST exact gauge fixing can be added to the action, even though it depends explicitly on the spacetime metric. The reason is that, since the gauge fixing is BRST exact, physical observables do not depend on the metric. In addition, the gauge fixing term provides quadratic terms for $\omega$, making a perturbative analysis possible.

Another important property of the model is the existence of a rich set of Ward identities, which are broken linearly, at most. This is a very welcome property which ensures their validity at quantum level \cite{Piguet:1995er}. In particular, the vierbein equation \eqref{vierbein1} ensures that the vierbein should not appear at the counterterm. This last feature is quite strong and ensures that the constraint could be employed at quantum level while maintaining $e$ classical.

To understand what a quantum version of the model would mean, let us consider a gauge fixed action $\widetilde{\Sigma}=\Sigma+s\Delta_{gf}$, enjoying the above discussed properties. The partition function can be written as
\begin{equation}
\widetilde{Z}=\int D\Phi \exp\{i\widetilde{\Sigma}(\Phi,e)\}\;,\label{z1}
\end{equation}
where
$D\Phi\equiv D\omega Dc D\bar{\sigma}D\sigma D\bar{\eta}D\eta D\bar{\theta}D\theta D\bar{\rho}D\rho D\bar{\gamma}D\gamma D\bar{c}Db$ while $\bar{c}$ is the Faddeev-Popov anti-ghost field and $b$ the Lautrup-Nakanishi field enforcing the referred gauge fixing. The partition function $\widetilde{\Sigma}$ defines the quantum version of the model coupled to the vierbein classical field. Since there is no previous dynamics for the spacetime, it is the model itself that defines the spacetime dynamics. Moreover, the classical limit of such model would provide a classical gravity limit, which is exactly the Lovelock-Cartan action \eqref{action1}. Hence, we have constructed a topological and BRST exact quantum model that addresses dynamics to spacetime and has the Lovelock-Cartan as its classical limit. In this limit, the identification of $\omega$ with the spin-connection and $\mu$ with the Newton's constant is natural.

\subsection{Further symmetry aspects}

Let us assume the existence of consistent BRST exact gauge fixed action which allows the construction of a suitable partition function $Z$ and the usual perturbative tools \cite{Baulieu:1988xs}. Moreover, it is also reasonable to assume that the gauge fixing would not spoil any of the Ward identities\footnote{See previous section.}. In addition, since the field equations \eqref{FE1} are exact or linearly broken, the validity of constraints \eqref{const1} at quantum level is ensured. 

One effect of the constraint \eqref{action6} is that, besides the BRST symmetry is preserved, its decomposition \eqref{rel0} is not. In fact, it is easy to check that
\begin{align}
s_o\Sigma=s_oS_c&=\Delta_b\;,\nonumber\\
\widetilde{\delta}\Sigma=\widetilde{\delta}S_c&=-\Delta_b\;,\label{so0}
\end{align}
which is consistent with the relations \eqref{rel0} and $s\Sigma=0$. As a consequence, we have the relations
\begin{eqnarray}
sZ&=&0\;,\nonumber\\
s_oZ&\neq&0\;,\nonumber\\
\widetilde{\delta}Z&\neq&0\;.\label{z0}
\end{eqnarray}
The field equations \eqref{FE1} can be written in the form of expectation values\footnote{The expectation values are taken with respect to the functional measure $D\Phi$ as defined in Sec. \ref{DISa}.} as
\begin{eqnarray}
\langle\sigma^a\rangle&=&m e^a\;,\nonumber\\
\langle\bar{\sigma}^a\rangle&=&m e^a\;,\nonumber\\
\langle\eta^a\rangle&=&0\;,\nonumber\\
\langle\bar{\eta}^a\rangle&=&0\;,\nonumber\\
-\langle\bar{\sigma}^a\rangle+\langle c\bar{\eta}^a\rangle&=&\langle s\bar{\eta}^a\rangle\;,\nonumber\\
\langle \eta^a\rangle-\langle c\sigma^a\rangle&=&\langle s\sigma^a\rangle\;,\nonumber\\
-\langle c\eta^a\rangle&=&\langle s\eta^a\rangle\;,\nonumber\\
-\langle c\bar{\sigma}^a\rangle&=&\langle s\bar{\sigma}^a\rangle\;.\label{ssb0}
\end{eqnarray}
Due to \eqref{z0}, the BRST operator $s$ commute with the expectation values in \eqref{ssb0}. Hence, due to the first four relations in \eqref{ssb0} and the $s$ invariance of the vierbein we have that $\langle s\bar{\eta}^a\rangle=\langle s\eta^a\rangle=\langle s\bar{\sigma}^a\rangle=\langle s\sigma^a\rangle=0$.
Moreover, from \eqref{brst1a} and \eqref{brs2a}, we have
\begin{eqnarray}
\langle s_o\bar{\eta}^a\rangle&=&-\langle c^a_{\phantom{a}b}\bar{\eta}^b\rangle\;,\nonumber\\
\langle s_o\bar{\sigma}^a\rangle&=&-\langle c^a_{\phantom{a}b}\bar{\sigma}^b\rangle\;,\nonumber\\
\langle s_o\sigma^a\rangle&=&-\langle c^a_{\phantom{a}b}\sigma^b\rangle\;,\nonumber\\
\langle s_o\eta^a\rangle&=&-\langle c^a_{\phantom{a}b}\eta^b\rangle\;,\label{brst1b}
\end{eqnarray}
and
\begin{eqnarray}
\langle\widetilde{\delta}\bar{\eta}^a\rangle&=&\langle\bar{\sigma}^a\rangle\;,\nonumber\\
\langle\widetilde{\delta}\bar{\sigma}^a\rangle&=&0\;,\nonumber\\
\langle\widetilde{\delta}\sigma^a\rangle&=&\langle\eta^a\rangle\;,\nonumber\\
\langle\widetilde{\delta}\eta^a\rangle&=&0\;.\label{brs2b}
\end{eqnarray}
 
Now, combining \eqref{ssb0}, \eqref{brst1b} and \eqref{brs2b}, we get
\begin{eqnarray}
\langle s_o\bar{\eta}^a\rangle&=&-m  e^a\;,\nonumber\\
\langle s_o\bar{\sigma}^a\rangle&=&0\;,\nonumber\\
\langle s_o\sigma^a\rangle&=&0\;,\nonumber\\
\langle s_o\eta^a\rangle&=&0\;,\label{brst1c}
\end{eqnarray}
and
\begin{eqnarray}
\langle\widetilde{\delta}\bar{\eta}^a\rangle&=&m e^a\;,\nonumber\\
\langle\widetilde{\delta}\bar{\sigma}^a\rangle&=&0\;,\nonumber\\
\langle\widetilde{\delta}\sigma^a\rangle&=&0\;,\nonumber\\
\langle\widetilde{\delta}\eta^a\rangle&=&0\;.\label{brs2c}
\end{eqnarray}
From \eqref{brst1c} and \eqref{brs2c} we understand that the breaking of the symmetries $\delta$ and $s_o$ compensate each other, as in \eqref{so0}, while the $s$ symmetry remains a symmetry of the model. In addition, it is clear that these breaks are directly related with the vierbein and the mass parameter since $s_o$- and $\widetilde{\delta}$-exact terms attain a non-vanishing vacuum expectation value equal to $m e^a$.

\section{Conclusions}\label{CONC}

We have constructed a massless gauge theory coupled with the vierbein field through algebraic constraints quadratic in the fields. Essentially, the action is composed by a topological and a BRST exact term. The constraints also carry a mass parameter which, eventually, is identified with Newton's constant. The interpretation of the model is that the gauge theory induces a dynamics for the spacetime, resulting in the  Lovelock-Cartan action \cite{Mardones:1990qc}. 

The constraints, being quadratic in the fields, ensure the validity of a rich set of symmetries. These symmetries, in the form of Ward identities, motivates the construction of a quantum version of the model. However, due to intricacies such as the gauge fixing problem and quantum stability, the formal analysis of the quantization of the model is left for a future paper.

Another possibility to be investigated is the generalization of the model to other dimensions, at least at classical level.

Finally, an extra remark is that, at classical level, the model can be simplified to consider only Lovelock gravity \cite{Lovelock:1971yv} or even general relativity. However, in a quantum version of such simplified models, it seems that the Ward identities are not strong enough to block the other terms of the Lovelock-Cartan. So, they would probably appear in the counterterm requiring their introduction in the bare action.

\section*{Acknowledgments}

The Conselho Nacional de Desenvolvimento Cient\'{i}fico e Tecnol\'{o}gico (CNPq-Brazil) and the Coordena\c c\~ao de Aperfei\c coamento de Pessoal de N\'ivel Superior (CAPES) are acknowledge for financial support.

\bibliographystyle{utphys2}
\bibliography{library}

\end{document}